\documentstyle[prl,aps,preprint,epsf]{revtex}   

\begin{document}

\title{The Origin of the Designability of Protein Structures} 

\author{Rie Tatsumi and George Chikenji}

\address{ Department of Physics,
          Graduate School of Science, Osaka University,\\
          Machikaneyama-cho 1-1, Toyonaka, Osaka 560-0043, Japan\\
        }

\date{\today}

\maketitle

\begin{abstract}

  We examined what determines the designability of 2-letter codes (H and P) lattice proteins from three points of view.
  First, whether the native structure is searched within
  all possible structures or within maximally compact structures.
  Second, whether the structure of the used lattice is bipartite or not.
  Third, the effect of the length of the chain, namely,
  the number of monomers on the chain. 
 We found that the bipartiteness of the lattice structure is not a main factor which determines the designability. 
Our results suggest that highly designable structures will be found when the length of the chain is sufficiently long to make the hydrophobic core consisting of enough number of monomers.

\end{abstract}




\section*{Introduction}
\label{sec:intro}

Natural proteins fold into unique compact structures in spite of 
the huge number of possible conformations\cite{Creighton:92}.  For most single domain proteins, each of these native structures corresponds to the global minimum of the free energy\cite{Anfinsen:73}.

It has been proposed phenomenologically that the number of possible structures of natural proteins is only about one thousand\cite{Chothia:92}, which suggests that many sequences can fold into one preferred structure. There have been theoretical studies for the existence of such preferred structures
\cite{Li:Helling:Tang:Wingreen:96,Irback:Sandelin:98,Li:Tang:Wingreen:98,Ejtehadi:Hamedani:Seyed-Allaei:Shahrezaei:Yahyanejad:97,Melin:Li:Wingreen:Tang:98}.

In many of theoretical studies for the protein folding, a simplified model called HP model 
\cite{Li:Helling:Tang:Wingreen:96,Irback:Sandelin:98,Ejtehadi:Hamedani:Seyed-Allaei:Shahrezaei:Yahyanejad:97,Melin:Li:Wingreen:Tang:98,Lau:Dill:89,Seno:Vendruscolo:Maritan:Banavar:96,Deutsch:Kurosky:96,Camacho:Thirumalai:93}
is adopted. 
HP model is one of 2-letter codes lattice models where a protein is represented by a self-avoiding chain of beads placed on a lattice, with two types of beads, hydrophobic(H) and polar(P). 
In the HP model, the energy of a structure is given by the nearest-neighbor topological contact interactions as
\begin{equation}
 H = - \sum_{i<j} E_{\sigma_i\sigma_j} \Delta(r_{i} - r_{j}) \label{HPhamil}
\end{equation}
where $i$ and $j$ are monomer indexes, $\{\sigma_i\}$ are monomer types ($\sigma=$  H or P);  $ \Delta(r_{i} - r_{j}) = 1 $ if $r_{i}$ and $r_{j}$ are topological nearest neighbors not along the sequence, and $ \Delta(r_{i} - r_{j}) = 0 $ otherwise.

Based on the HP model, a concept of {\em designability} has recently been introduced\cite{Li:Helling:Tang:Wingreen:96}; the number of sequences that have a given structure as their non-degenerate ground state (native state) is called  the {\em designability} of this structure. 
When many sequences have a common native structure, one say that the structure is {\em highly designable}.  Adding to the importance in the protein design problem, the designability also have evolutional significance because highly designable structures are found to be relatively stable against mutations. 

In the original study of H. Li {\em et al\/}.\cite{Li:Helling:Tang:Wingreen:96}, HP models on the square and cubic lattices are employed, with the energy parameters in eq.(\ref{HPhamil}) being $ E_{HH} = -2.3,$ $ E_{HP} = -1.0,$ $ E_{PP} = 0.0 $.  For each sequence, they calculated the energy over all maximally compact structures and picked up the native structure.  The results indicated that highly designable structures actually exist on both lattices.

A. Irb\"ack and E. Sandelin studied the HP models on the square and triangular lattices\cite{Irback:Sandelin:98}.  They adopted different energy parameters from H. Li {\em et al\/}. \cite{Li:Helling:Tang:Wingreen:96}, namely,  $ E_{HH} = -1,$ $ E_{HP} = E_{PP} = 0 $.  In the calculation of the designability, they considered all the possible structures, not restricting to the maximally compact ones.  For the square lattice, they confirmed the existence of the highly designable structures as in Ref.\cite{Li:Helling:Tang:Wingreen:96}.  For the triangular lattice, however, no such structures were found. In addition to the nearest-neighbor topological contact interactions, they considered local interactions represented by the bend angle and calculated the designability. Indeed the local interactions reduced degeneracy ({\it i.e.}, the number of sequences which have non-degenerate ground state increased) and made the designability higher. But they found that the designability on the square lattice was still much higher than that on the triangular lattice. They concluded that the difference in the designability for these two lattices are related to the even-odd problem, that is, whether the lattice structure is bipartite or not.

Quite recently, H. Li {\em et al\/}. proposed a new model based on the HP model on the square lattice \cite{Li:Tang:Wingreen:98}.  In the model, the hydrophobic interaction is treated in such a way that the energy decreases if the hydrophobic residue is buried in the core.  They justify this treatment in two reasons:
(1) the hydrophobic force which is dominant in folding
\cite{Dill:90,Dill:Bromberg:Yue:Fiebig:Yee:Thomas:Chan:95} originates from aversion of hydrophobic residues from water.
(2) the Miyazawa-Jernigan matrix \cite{Miyazawa:Jernigan:85} 
contains a dominant hydrophobic interaction of the linear form
$ E_{\alpha\beta} = h_\alpha + h_\beta $
\cite{Li:Tang:Wingreen:97}.
They took
\begin{equation}
 H = - \sum_{i=1}^N s_{i} h_{i}  \label{newHP}
\end{equation}
where $\{h_i\}$ represent a sequence : $ h_i = 1 $ if the {\it i}-th amino acid is H-type and $ h_i = 0 $ if it is P-type.  And $\{s_i\}$ represent a structure : $ s_i = 0 $ if the {\it i}-th amino acid is on the surface and $ s_i = 1 $ if it is in the core.  They calculated the designability over all maximally compact structures, whose result is consistent with their former study\cite{Li:Helling:Tang:Wingreen:96} [See Table.~\ref{table1}~].

In our view, there are many points to be explored further for the designability problem.  First, since the structures of natural proteins are compact but not necessarily ``maximally compact'' in general, how can we justify the discussion where only the maximally compact structures are taken into account? 
 Second, is it adequate to consider only nearest-neighbor interactions?  Properties of a system with only nearest-neighbor interactions are directly affected by the lattice structure, in particular, whether the lattice is bipartite or not.  Is it good, only from these facts, to conclude immediately that the absence of the highly designable structures on the triangular lattice should be ascribed to the even-odd problem associated with the triangular lattice?\cite{Irback:Sandelin:98}  One should discuss the problem on the triangular lattice by using a model like the one in Ref.\cite{Li:Tang:Wingreen:98} where the interactions do not depend on the contact between monomers, hence, {\em do not directly reflect the non-bipartiteness}.

Our aim of this paper is to examine the above points
 and clarify what determines the designability of protein structures.
  We use a new model with a 2-letter codes (H and P) on the square and triangular lattices and calculate the designability over {\em all possible} structures.
 In our model, based on Ref.\cite{Li:Tang:Wingreen:98}, the energy increases if the hydrophobic residue is exposed to the solvent.
We will call this model ``solvation model''.
In brief, the solvation model is a 2-letter codes lattice model where the hydrophobic force to form a core is dominant and the interactions do not directly reflect the bipartiteness. Using the solvation model and the HP model, we investigate model-independent properties of designability.


\section*{The Models}
\label{sec:models}

In the solvation model, based on Ref.
\cite{Li:Tang:Wingreen:98}, a protein is represented by a self-avoiding chain of beads with two types H and P, placed on a lattice.
A sequence is specified by a choice of monomer types at each position on the chain.

We used two-dimensional lattice models because a computable length by numerical enumeration of the full conformational space is limited (square lattice : 18, triangular and cubic lattices : 13).  Even with this chain-length limitation, we can make a ``hydrophobic core'' in two dimensions, in contrast with the three-dimensional case.

A structure is specified by a set of coordinates for all the monomers and is mapped into the number of contacts with the solvent.  In our model, the total energy is given in terms of the monomer-solvent interactions, and depends only on the number of contacts with the solvent:
\begin{equation}
 H = \sum_{i=1}^N E_{s_i} h_i  \label{OurModel}
\end{equation}
where $ \{h_i\} $ represent a sequence : $ h_i = 1 $ if the {\it i}-th monomer is the H-type and $ h_i = 0 $ if it is P-type.  The variable $ s_i $ denotes the number of contacts with the solvent, for example, $ s_i = \{ 0, 1, 2, 3\} $ on the square lattice and $ s_i = \{ 0, 1, 2, 3, 4, 5\} $ on the triangular lattice.  In other words, $ s_i = 0 $ means that the $i$-th monomer is buried away from the solvent.  We take $ E_0 = 0,$ $ E_1 = \sqrt{2},$ $ E_2 = \sqrt{7},$ $ E_3 = \sqrt{13},$ $ E_4 = \sqrt{19},$ $ E_5 = \sqrt{23} $.  That is, the possible minimum energy is zero. 
And these parameters are selected so that the larger the number of contacts with the solvent is, the more the degree of energy increase is; the hydrophobic residue is energetically unfavorable to be at the corner\cite{Yue:Dill:95,Yue:Dill:93}.
Although the choice of these values is somewhat arbitrary, we have considered the following points: (1) these values should not increase too much rapidly with the increase in the number of contacts with the solvent, 
and (2) the way of choosing these values must not bring about nonessential accidental degeneracies (due to simple rational ratios between the parameters) 
\cite{Ejtehadi:Hamedani:Shahrezaei:98}. 

Using the model on the square and triangular lattices, we calculate the designability for all the $ 2^N $ sequences, where $N$ is the number of monomers, by exact computer-enumeration method over the full conformational space.  To get correct data, we exclude overcounting coming from redundant structures which are mutually related by rotation, reflection and reverse-labeling.

On the basis of data obtained by the solvation model and the HP model, we examine what determines the designability from three points of view: (1) the effect of the search-space restriction, namely, the search within maximally compact structures (in this paper, we just used maximally compact structures as a simplest example of the search-space restriction, and we may consider other one, {\it e.g.}, structures with the biggest core), (2) the effect of the lattice structure, namely, whether the lattice is bipartite or not (or, equivalently, the even-odd problem), (3) the effect of the number of monomers (or, the length of the chain).

\section*{Results and Discussion}
\label{sec:results and discussion}

Let us now give results of calculations.

\hspace*{-7mm}(1) The effect of the search within maximally compact structures

\begin{figure}
\begin{center}
\leavevmode
\epsfxsize=135mm
\epsfbox{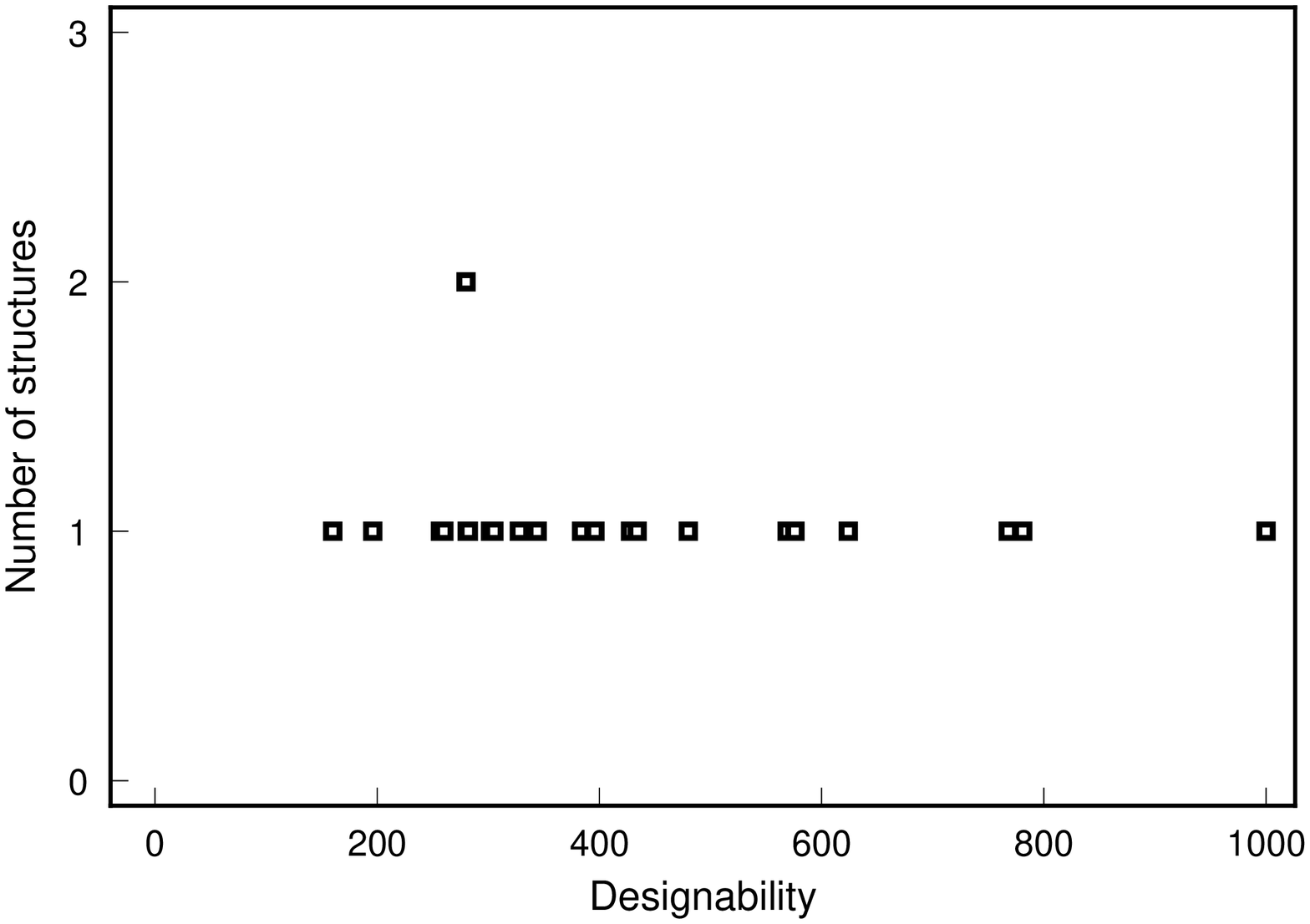}
\end{center}
\caption{The designability calculated over maximally compact structures on the square lattice for N=16. The term of ``Number of structures'' at the vertical axis means how many structures with the ``Designability'' at the horizontal axis there are.}
\label{fig1} 
\end{figure}

\begin{figure}
\begin{center}
\leavevmode
\epsfxsize=135mm
\epsfbox{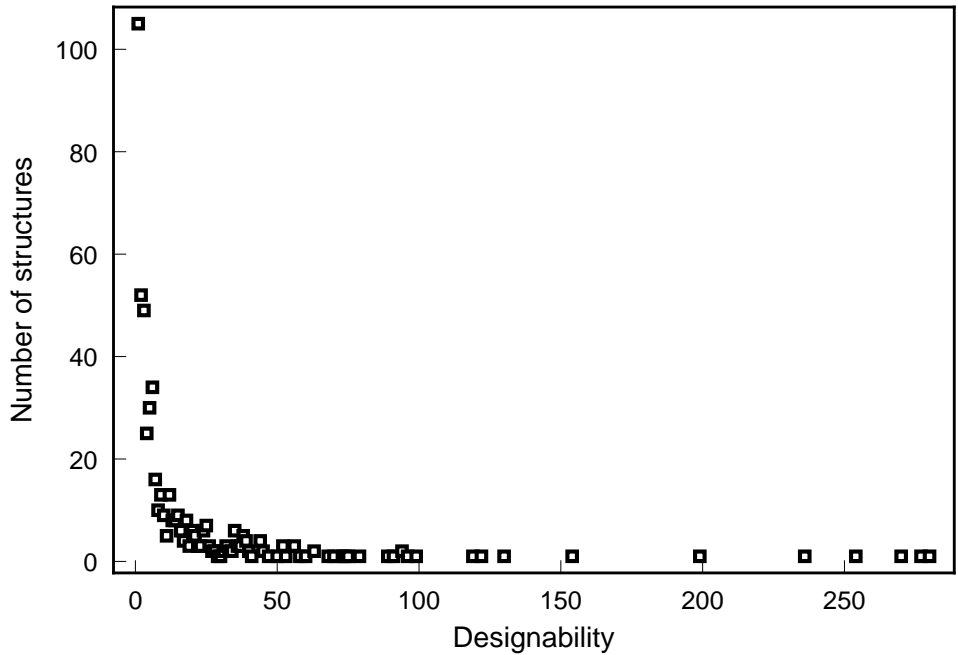}
\end{center}
\caption{The designability calculated over all possible structures on the square lattice for N=16.}
\label{fig2}
\end{figure}

In Fig.~\ref{fig1}, we show the designability calculated on the square lattice for $N=16$, using maximally compact structures.  For comparison, in Fig.~\ref{fig2}, we show the designability of the same system without the search-space restriction ({\it i.e.}, search over all possible structures).  In both cases, there are some highly designable structures.  However, these structures are not common to both cases.  In Fig.~\ref{fig2}, the number of sequences that have native structures is 8277, but the number of sequences that have maximally compact structures as native is only 1087 out of 8277.
  That is, most sequences that have native structures have non-maximally compact structures as native. 
The importance of non-maximally compact structures has also been pointed out for the HP model
\cite{Irback:Sandelin:98,Yue:Fiebig:Thomas:Chan:Shakhnovich:Dill:95,Frauenkron:Bastolla:Gerstner:Grassberger:Nadler:98,Vendruscolo:Maritan:Banavar:97,Klimov:Thirumalai:96}.
 These facts imply that it is not good to calculate the designability over only maximally compact structures.
  Such calculation picking up a ``native'' structure out of maximally compact structures, is not correct if the true native structure is non-maximally compact.  
Further, when the lowest-energy non-maximally compact structure and the lowest-energy maximally compact structure are degenerate, there is no native structure (native structure must be non-degenerate),  but the restricted-search-space calculation gives a false result that there is a native (and maximally compact) structure.  We should say that the designability calculated over only maximally compact structures may be erroneous.

\hspace*{-7mm}(2) The effect of the lattice structure: bipartite or non-bipartite

In two previous studies using the HP model \cite{Li:Helling:Tang:Wingreen:96,Irback:Sandelin:98}, interactions of the system directly reflected whether the lattice is bipartite or not. Moreover the designability on the triangular lattice was calculated with the energy parameters in eq.(\ref{HPhamil}) being $ E_{HH} = -1, E_{HP} = E_{PP} = 0 $, which would cause accidental degeneracies. In their results, highly designable structures were not found for the triangular lattice.  Also, it seemed that native structures are likely to contain the hydrophobic core where a group of hydrophobic monomers contact with each other; such contact can be made only if the distance between the monomers along the sequence is odd. Therefore, the bipartiteness has been thought to be a main source of the designability.\cite{Li:Helling:Tang:Wingreen:96,Irback:Sandelin:98,Shakhnovich:Abkevich:Ptitsyn:96}. If so, highly designable structures do not actually exist, 
{\it i.e.}, the concept of {\it designability} itself could be meaningless.
On the other hand, if such preferred structures should exist on the basis of the proposal by C. Chothia
\cite{Chothia:92}, the use of the lattice model would be inadequate.
Then, we used the solvation model, which does not directly reflect the bipartiteness, and calculated the designability on the square and triangular lattices.
Besides, we also calculated the designability on the triangular lattice using the HP model, with the energy parameters being $ E_{HH} = -2.3, E_{HP} = -1.0, E_{PP} = 0.0 $.

In Table.~\ref{table2}, we show the total number of sequences that have non-degenerate ground state ($S_n$) and the highest designabilities ($D_h$) on the triangular lattice for $N=13$, obtained by using different interactions. 
This result shows that, even if we take different values of energy parameters, or even if we use the solvation model, the triangular lattice is still unfavorable for the designability although $S_n$ varies largely.  On the other hand, for the square lattice, highly designable structures are found in the solvation model as well as in the HP model (Fig.~\ref{fig2}). 
 These results imply that the absence of the highly designable structures for the triangular lattice should not be ascribed to the even-odd problem (or, the non-bipartiteness), but to other reasons.
The properties that highly designable structures are found on the square lattice and no such structures are found on the triangular lattice might be general in 2-letter codes lattice models where the hydrophobic force is dominant.

\hspace*{-7mm}(3) The effect of the number of monomers

Then, why are the highly designable structures absent for the triangular lattice?  Smallness of number of monomers (in other words, the length of a chain is too short), may be a possible reason.  Important object in the protein structure is the hydrophobic core which consists of  buried monomers in no contact with the solvent.  Recall that the limit of a computable length by exact enumeration of the full conformational space on the triangular lattice is 13.  The biggest core which we can make by using this limited length is the one which consists of only three monomers; the length is too short for the hydrophobic force to form a core.
This monomer-number effect is also found on the square lattice.  Consider the following conditions: at least ten sequences have a given structure as their native state, and at the same time, there are at least five such structures. Only if these conditions are satisfied, let us say that ``there are highly designable structures.'' Then, at $N=10$ or less, there are no highly designable structures even for the square lattice [Table.~\ref{table3}, Table.~\ref{table4}].  This result implies that, when we discuss whether there are highly designable structures or not, we need a long chain enough to make a core of enough size.  This further implies that, in three-dimensional case, we will need a chain of longer length than that in two-dimensional case to make a core. 

Let us see Table.~\ref{table3}, Table.~\ref{table4} and Table.~\ref{table5}. In Table.~\ref{table5}, we show the designability calculated on the triangular lattice for $N=13$. 
On the square lattice for $N=10$, the biggest core consists of two monomers.
Both on the triangular lattice for $N=13$ and on the square lattice for $N=11$, the biggest core consists of three monomers. We see that the triangular lattice is unfavorable for designability compared with square lattice, even when the biggest possible core size is same or a little larger.
A possible reason would be the number of all possible structures, particularly the number of structures with the biggest core. As the length of a chain becomes long, the number of all possible structures increases almost exponentially as $\mu^N$ ( $2<\mu<3$ for the square lattice, and $4<\mu<5$ for the triangular lattice)\cite{Madras:Slade}.
On the triangular lattice for $N=13$, the number of all possible structures is 6,279,601 and the number of structures with the biggest core is 4,110 out of them. On the other hand, on the square lattice for $N=10, 11$, the number of all possible structures is 2,034, 5,513 and the number of structures with the biggest core is 23, 5, respectively.
Thus the number of all possible structures and the number of structures with the biggest core on the triangular lattice are much larger than those on the square lattice
\cite{Vendruscolo:Subramanian:Kanter:Domany:Lebowitz:99}.
In consequence, the degeneracy tends to grow, which is unfavorable for designability. In this view, designable structures on the triangular lattice would be more difficult to appear than on the square lattice.

\section*{Summary}
\label{sec:summary}

We have calculated the designability of the protein structure using the solvation model and the HP model, to deduce model-independent properties of designability. 
The solvation model introduced in this paper satisfies two conditions: (1)the hydrophobic force is dominant, (2) the model does not directly reflect the bipartiteness. We have examined what determines the designability from three points of view: effect of restricted search within maximally compact structures, the bipartite/non-bipartite effect, the length of the chain.

In result, we have found that it is inadequate to calculate the designability within maximally compact structures. Our results imply that the reason why no highly designable structures on the triangular lattice have been found is not the non-bipartiteness.
We suppose that the main factor which affects the designability is the chain length, because for sufficiently large hydrophobic core to form, long enough chains are required. 
Triangular lattice is more unfavorable for the designability than square lattice irrespective of models or energy parameters, probably because the number of all possible structures is large. 
However, if we can deal with longer chain than in the present study, it is possible that we find highly designable structures even on the triangular lattice.
The calculations of the designability for longer chains on the triangular lattice are highly desirable.
These conclusions would apply to a wide variety of 2-letter codes lattice models, where the hydrophobic force is dominant, regardless of energy parameters and further details of the model.

Though a concept of designability is currently defined for a 2-letter codes lattice model, our final goal is to examine whether natural proteins have highly designable structures.
Therefore it is an interesting problem to extend the study of the designability for a 20-letter codes model ({\it e.g.}, MJ model\cite{Miyazawa:Jernigan:85}, KGS model\cite{Kolinski:Godzik:Skolnick:93}) and an off-lattice model.
Substituting 20-letter codes for 2-letter codes certainly reduces degeneracy, and most of all sequences come to have a structure as non-degenerate ground state ({\it i.e.}, native structure).

\section*{Acknowledgements}
\label{sec:acknowledgements}

We would like to thank Y. Akutsu and M. Kikuchi for useful discussions and careful reading of the manuscript.

\begin{table}
\begin{center}
\small
\begin{tabular}{|c|ccc|} \hline 
               & H. Li {\em et al\/}.\cite{Li:Helling:Tang:Wingreen:96}
& A. Irb\"ack and E. Sandelin\cite{Irback:Sandelin:98} 
& H. Li {\em et al\/}.\cite{Li:Tang:Wingreen:98} \\ \hline
lattice & square and cubic & square and triangular & square \\
\smash{\lower3.0ex\hbox{interaction}} & \multicolumn{2}{c}{\smash{\lower3.0ex\hbox{nearest-neighbor}}} & depend on   \\
                 &             &         & the position of a H \\
Hamiltonian & \multicolumn{2}{c}{$ H = - \sum_{i<j} E_{\sigma_i\sigma_j} \Delta(r_{i} - r_{j})$}  & $ H = - \sum_{i=1}^N s_{i} h_{i} $ \\ 
energy parameter & \smash{\lower3.0ex\hbox{(-2.3, -1.0, 0.0)}} &  \smash{\lower3.0ex\hbox{(-1, 0, 0)}} & \\
$(E_{HH}, E_{HP}, E_{PP})$ &  &  & \\
conformational space & maximally compact & all  & maximally compact  \\
highly designable & found on both lattices & found on square lattice & found \\
structures &  & but not found on triangular one & \\
\hline 
\end{tabular}
\end{center}
\caption{The difference among three researches is showed.
Each variable in the Hamiltonian is defined in the text.} 
\label{table1}
\end{table}

\begin{minipage}{6.5cm}
\begin{table}
\begin{center}
\begin{tabular}{|c|cc|} \hline
                         & $ S_n $ & $ D_h $ \\
\hline
HP model (-1, 0, 0) & 0 & 0 \\
HP model (-2.3, -1.0, 0.0) & 129 & 3 \\ 
solvation model & 7  & 1 \\
\hline
\end{tabular}
\end{center}
\caption{$ S_n $ and $ D_h $ on the triangular lattice for $N=13$. The parenthesis corresponds to energy parameters $ (E_{HH}, E_{HP}, E_{PP}) $. The data in the HP model with the energy parameters being $ E_{HH} = -1, E_{HP} = E_{PP} = 0 $ was obtained by A. Irb\"ack and E. Sandelin
[5].
$ S_n $ and $ D_h $ are defined in the text.}
\label{table2}
\end{table}
\end{minipage}

\begin{minipage}{8.cm}
\begin{table}
\begin{center}
\begin{tabular}{|cc|} \hline
Designability & Number of Structures \\
\hline
1 & 1 \\
2  & 2 \\
3  & 4 \\
4  & 3 \\
5  & 4 \\
6  & 1 \\
10 & 2 \\
12 & 1 \\
\hline
\end{tabular}
\end{center}
\caption{The designability calculated over all possible structures on the square lattice for $N=10$. The term of ``Number of Structures'' means how many structures with the ``Designability'' on its left there are.}
\label{table3}
\end{table}
\end{minipage}

\begin{minipage}{8.cm}
\begin{table}
\begin{center}
\begin{tabular}{|cc|} \hline
Designability & Number of Structures \\
\hline
1 & 5 \\
2  & 11 \\
3  & 4 \\
4  & 1 \\
5  & 3 \\
8  & 1 \\
10 & 1 \\
13 & 1 \\
18 & 1 \\
29 & 1 \\
36 & 1 \\
43 & 1 \\
\hline
\end{tabular}
\end{center}
\caption{The designability calculated over all possible structures on the square lattice for $N=11$.}
\label{table4}
\end{table}
\end{minipage}

\begin{minipage}{8.cm}
\begin{table}
\begin{center}
\begin{tabular}{|cc|} \hline
Designability & Number of Structures \\
\hline
1 & 7  \\
\hline
\end{tabular}
\end{center}
\caption{The designability calculated over all possible structures on the triangular lattice for $N=13$.}
\label{table5}
\end{table}
\end{minipage}



\begin{references}
  
\bibitem{Creighton:92} T.E. Creighton (ed.), Protein Folding (Freeman,
  New York, 1992)

\bibitem{Anfinsen:73}
C. Anfinsen, Science {\bf 181} 223 (1973)

\bibitem{Chothia:92}
C. Chothia, Nature {\bf 357} 543 (1992)

\bibitem{Li:Helling:Tang:Wingreen:96}
H. Li, R. Helling, C. Tang and N. Wingreen, Science {\bf 273} 666 (1996)

\bibitem{Irback:Sandelin:98}
A. Irb\"ack and E. Sandelin, J. Chem. Phys. {\bf 108}, 2245 (1998)

\bibitem{Li:Tang:Wingreen:98}
H. Li, C. Tang, and N. S. Wingreen, Proc. Natl. Acad. Sci. USA {\bf 95}, 4987 
(1998)

\bibitem{Ejtehadi:Hamedani:Seyed-Allaei:Shahrezaei:Yahyanejad:97}
M. R. Ejtehadi, N. Hamedani, H. Seyed-Allaei, V. Shahrezaei and M. Yahyanejad,
J. Phys. A {\bf 29}, 6141 (1998)

\bibitem{Melin:Li:Wingreen:Tang:98}
R. Melin, H. Li, N. S. Wingreen and C. Tang, cond-mat/9806197 preprint.

\bibitem{Lau:Dill:89}
K. F. Lau and K. A. Dill, Macromolecules {\bf 22} 3986 (1989)

\bibitem{Seno:Vendruscolo:Maritan:Banavar:96} 
F. Seno, M. Vendruscolo, A. Maritan and J. R. Banavar, PRL {\bf 77}, 1901 (1996)  

\bibitem{Deutsch:Kurosky:96} 
J. M. Deutsch and T. Kurosky, PRL {\bf 76}, 323 (1996) 

\bibitem{Camacho:Thirumalai:93} 
C. J. Camacho and D. Thirumalai, PRL {\bf 71}, 2505 (1993)

\bibitem{Dill:90} K. A. Dill, Biochemistry {\bf 29}, 7133 (1990)

\bibitem{Dill:Bromberg:Yue:Fiebig:Yee:Thomas:Chan:95} 
K. A. Dill, S. Bromberg, K. Yue, K. M. Fiebig, D. P. Yee, P. D. Thomas and H. S. Chan, Protein Science {\bf 4}, 561 (1995)

\bibitem{Miyazawa:Jernigan:85}
S. Miyazawa and R. L. Jernigan, Macromolecules {\bf 18}, 534 (1985)

\bibitem{Li:Tang:Wingreen:97} 
H. Li, C. Tang and N. S. Wingreen, PRL {\bf 79}, 765 (1997)

\bibitem{Yue:Dill:95} 
K. Yue and K. A. Dill, Proc. Natl. Acad. Sci. USA {\bf 92}, 146 (1995)
  
\bibitem{Yue:Dill:93}
K. Yue and K. A. Dill, PRE {\bf 48}, 2267 (1993)

\bibitem{Ejtehadi:Hamedani:Shahrezaei:98}
M. R. Ejtehadi, N. Hamedani and V. Shahrezaei, cond-mat/9811127 preprint.

\bibitem{Yue:Fiebig:Thomas:Chan:Shakhnovich:Dill:95}
K. Yue, K. M. Fiebig, P. D. Thomas, H. S. Chan, E. I. Shakhnovich and K. A. Dill, Proc. Natl. Acad. Sci. USA {\bf 92}, 325 (1995)

\bibitem{Frauenkron:Bastolla:Gerstner:Grassberger:Nadler:98}
H. Frauenkron, U. Bastolla, E. Gerstner, P. Grassberger and W. Nadler, PRL {\bf 80}, 3149 (1998)

\bibitem{Vendruscolo:Maritan:Banavar:97}
M. Vendruscolo, A. Maritan and J. R. Banavar, PRL {\bf 78}, 3967 (1997)

\bibitem{Klimov:Thirumalai:96}
D. K. Klimov and D. Thirumalai, PRL {\bf 76}, 4070 (1996)

\bibitem{Shakhnovich:Abkevich:Ptitsyn:96}
E. Shakhnovich, V. Abkevich and O. Ptitsyn, Nature {\bf 379}, 96 (1996)

\bibitem{Madras:Slade}
N. Madras and G. Slade, The Self-Avoiding Walk (Birkh\"auser, Boston, 1993)

\bibitem{Vendruscolo:Subramanian:Kanter:Domany:Lebowitz:99}
M. Vendruscolo, B. Subramanian, I. Kanter, E. Domany and J. Lebowitz, 
PRE {\bf 59}, 977 (1999)
: In this paper, it is showed that the larger the number of all possible structures is, the larger the number of structures with the same number of contacts between monomers is.

\bibitem{Kolinski:Godzik:Skolnick:93}
A. Kolinski, A. Godzik and J. Skolnick, J. Chem. Phys. {\bf 98}, 7420 (1993)

\end{references}
\end{document}